\documentclass[doublecol]{epl2} 
\usepackage{amsmath}
\usepackage{amssymb}
\newcommand{\bv}{{\bf v}}

\title{The unbearable lightness of Restricted Boltzmann Machines: Theoretical Insights and Biological Applications}

\shorttitle{The unbearable lightness of Restricted Boltzmann Machines} 

\author{Giovanni di Sarra \inst{1} \and Barbara Bravi \inst{2} \and Yasser Roudi \inst{3}}
\shortauthor{G. di Sarra \etal}

\institute{                    
  \inst{1} Kavli Institute for Systems Neuroscience, NTNU- Olav Kyrres Gate 30, Trondheim, Norway\\
  \inst{2} Department of Mathematics, Imperial College London, London, SW7 2AZ, United Kingdom\\
  \inst{3} Department of Mathematics, King's College London-Strand, London, WC2R 2LS, United Kingdom
}

\abstract{Restricted Boltzmann Machines are simple yet powerful neural networks. They can be used for learning structure in data, and are used as a building block of more complex neural architectures. At the same time, their simplicity makes them easy to use, amenable to theoretical analysis, yielding interpretable models in applications. Here, we focus on reviewing the role that the activation functions, describing the input-output relationship of single neurons in RBM, play in the functionality of these models. We discuss recent theoretical results on the benefits and limitations of different activation functions. We also review applications to biological data analysis, namely neural data analysis, where RBM units are mostly taken to have sigmoid activation functions and binary units, to protein data analysis and immunology where non-binary units and non-sigmoid activation functions have recently been shown to yield important insights into the data. Finally, we discuss open problems addressing which can shed light on broader issues in neural network research.}

\begin{document}

\maketitle

\begin{figure}[h]
\centering
\includegraphics[width=8.8cm]{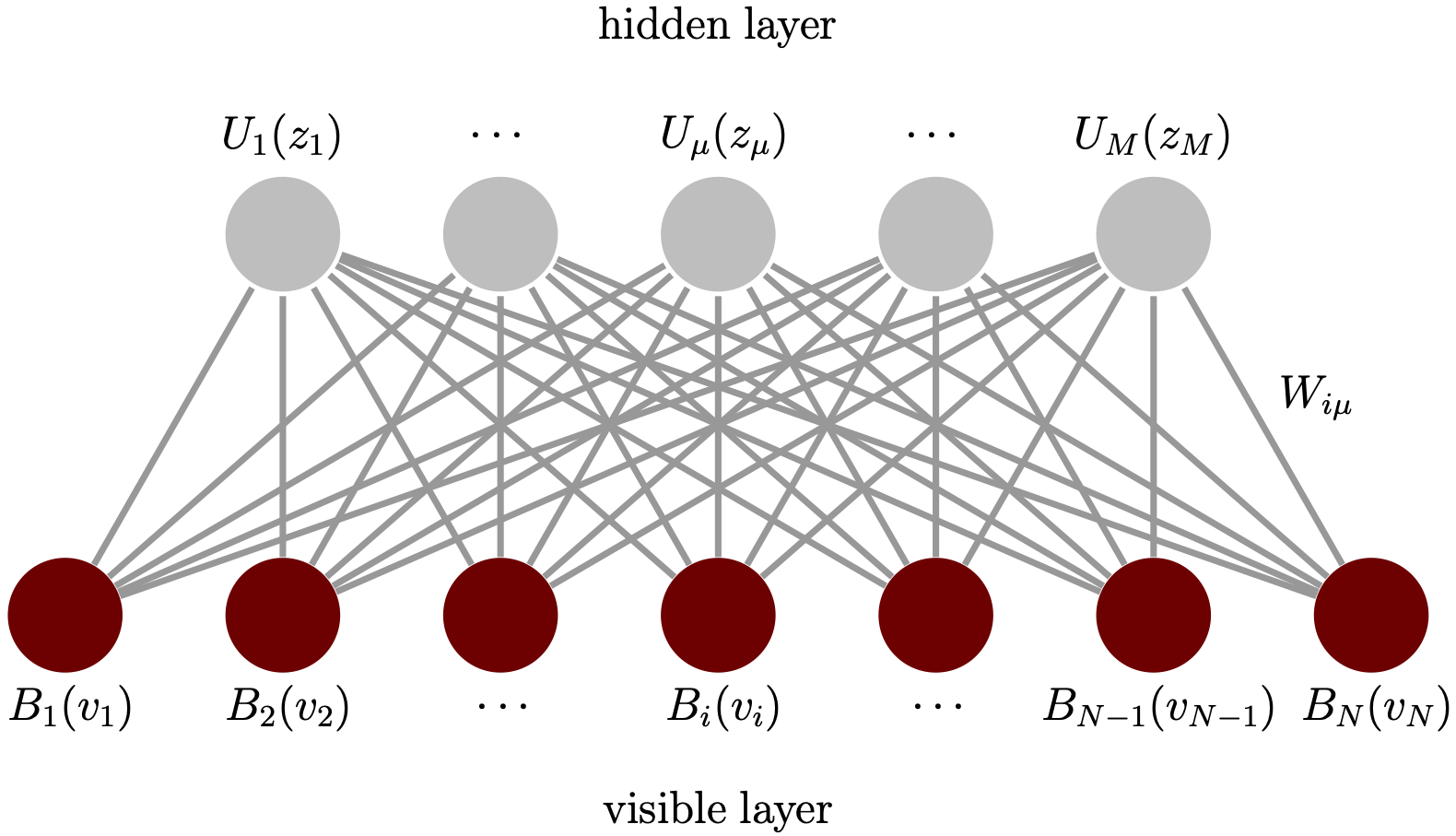}
\caption{RBM structure: two sets of variables $\{v_i\}_{i=1}^N$ and $\{z_{\mu}\}_{\mu =1}^M$ are organized into two layers, connected by weights $W_{i\mu}(v_i)$ living on a bipartite graph. Each set is subject to potentials $B_{i}$ and $U_{\mu}$ acting on single visible and hidden units, respectively.}
\label{fig:1}
\end{figure}

In this Perspective we discuss recent theoretical work on the properties of Restricted Boltzmann Machines (RBMs), focusing on how the activation function describing the input-output relationship of single neurons influence these properties. We also discuss how different choices of activation function make RBMs suitable for a range of applications to biological data.\\
A Restricted Boltzmann Machine is a two-layer stochastic neural network comprised of a set of visible (observed) units $\{v_i\}_{i=1}^N$, a set of hidden units $\{z_{\mu}\}_{\mu =1}^M$ and weights $W_{i\mu}$ connecting units in different sets; see Fig.\ \ref{fig:1}. The state of units in an RBM are determined via the Boltzmann distribution
\begin{subequations}
\begin{align}
\begin{split}
&p(\textbf{v},\textbf{z}) = \frac{1}{Z} e^{-\mathcal{H}(\textbf{v},\textbf{z})}
\end{split}\\
\begin{split}
&\mathcal{H}(\textbf{v},\textbf{z}) = -\sum_{i=1}^{N} B_{i}(v_{i}) - \sum_{i,\mu}^{N,M} z_{\mu} W_{i\mu}(v_{i}) -\sum_{\mu=1}^{M}U_{\mu}(z_{\mu}),
\end{split}
\end{align}
\label{eq:pdef}
\end{subequations}
where $W_{i\mu}(v_i)$ is often in the form of $w_{i\mu} v_i$ and $w_{i\mu}$ is the weight between the visible unit $i$ and the hidden unit $\mu$. However, as we will see later, other choices for $W_{i\mu}(v_i)$ are also used. When $W_{i\mu}(v_i)=0$, the states of observed and hidden units are determined by the potentials $B_{i}(\cdot)$ and $U_{\mu}(\cdot)$. The normalization factor, $Z$, is also called the partition function, in analogy to statistical mechanics. In what follows, we label visible elements with roman subscripts (e.g.~$v_{i}$, $B_{i}$), and hidden elements with Greek subscripts (e.g.~$z_{\mu}$, $U_{\mu}$).
\\
Although in the original formulation of RBMs, the units took few discrete states \cite{smolensky1986information}, continuous-valued or multi-state units are commonplace today \cite{bulso2021restricted, tubiana2019}. Here, unless otherwise stated, we assume binary visible units, $v_i \in \{0,1\}$, and linear visible potentials $B_i(v_i) = b_i v_i$, where $b_i \in \mathbb{R}$ are called visible biases. We do not, however, make such restrictions on the hidden units unless explicitly stated. 
The functions $U_{\mu}$ determine the form of the activation function and the set of states that hidden units can take. For example, if $U_{\mu}= c_{\mu} z_{\mu}$ for $z_{\mu} = 0,1$, and $U_{\mu}= +\infty$ otherwise, one recovers the standard RBM with binary units, $c_{\mu}$ being the biases acting on the hidden units. In this case, denoting the input to hidden units by $h_{\mu}\equiv c_{\mu}+\sum_{\mu} W_{i\mu}(v_{i})$ yields $p(z_{\mu}|h_{\mu})\propto \exp[z_{\mu}h_{\mu}]$ for $z_{\mu} = 0,1$, and $p(z_{\mu}|h_{\mu}) = 0$ otherwise. The mode of $p(z_{\mu}|h_{\mu})$ will be $\hat{z}_{\mu}=\Theta(h_{\mu})$, where $\Theta(\cdot )$ is the Heaviside function and the mean has a sigmoidal relationship to $h_{\mu}$. The functional form of dependence of this mode or mean on $h_{\mu}$ is what we denote as the activation function of the hidden units, in this case a step or a sigmoid function.\\
Although the state of the art deep neural networks architectures often surpass RBMs in performance, RBMs are still very useful, as the examples discussed below show, by balancing performance with being more amenable to a controlled inspection and interpretability. Furthermore, the tractability of RBMs combined with high representational capacity, makes them an ideal device to improve our current understanding of the dynamics and outcome of parameter learning from structured data. It can be shown that for binary hidden and visible units, RBMs can approximate any distribution arbitrarily well if the hidden layer is sufficiently large \cite{le2008representational,montufar}. In this sense, they are universal approximators.\\
The RBM is trained on data that are taken to be samples of the states of the visible nodes. For example, the data set could be a set of black and white images, in which case one assigns $v_i=0$ to the $i$th pixel if it is white and $v_i=1$ if it is black. The weights and potentials are then adjusted so that samples from the RBM, which are distributed according to $p(\bv)=\sum_{\mathbf{z}} p(\textbf{v},\textbf{z})$, resemble the distribution of the data. Such parameter learning can be done using gradient-based maximization of the likelihood \cite{ACKLEY1985147}, which involves the computation of derivatives of $Z$. Calculating these derivatives exactly becomes intractable when there are many units, but a variety of efficient approximate methods exist. Commonly used methods are Contrastive Divergence (CD) \cite{hinton2002training} and its variants Persistent Contrastive Divergence (PCD) \cite{tieleman2008training} and Tempered Contrastive Divergence (TCD) \cite{desjardins2010parallel} that rely on different Monte Carlo Markov Chain (MCMC) sampling methods; see \cite{fischer2014training} for a review of RBM training algorithms. Although updating parameters with CD yields increasing and large likelihood values, the resulting trained RBM may not act as a good generative model mainly due to the lack of convergence of the MCMC approach, requiring careful choices to be made depending on the specific goal of the training \cite{decelle2021equilibrium}.\\

{\bf Non-Sigmoid activation function.} Besides the standard sigmoid hidden activation function, choices of $U_{\mu}(z_\mu)$ leading to non-sigmoidal ones have also been analytically studied. An important case studied by Barra et al. \cite{barra2012equivalence} 
is that of $U_{\mu}(z_\mu) = z^2_{\mu}/2 + c_{\mu} z_{\mu}$ for $z_{\mu} \in \mathbb{R}$, which implies a linear activation function where $\hat{z}_{\mu} = h_{\mu}$. This choice yields the visible marginal distribution $p(\textbf{v})\propto \exp[\sum_{i<j} J_{ij} v_i v_j]$ with $J_{ij} = \sum_{\mu} w_{i \mu } w_{j \mu }$, resembling that of the Hopfield model with stored patterns $\xi_{i}^{\mu} = w_{ i \mu}$. When $M\ge N-1$, the linear hidden activation function can yield any pairwise distribution over the visible nodes. If $M<N-1$, it can yield a pairwise interaction matrix with minimum Frobenius distance to arbitrary visible pairwise interactions \cite{bulso2021restricted}. \\
For other choices of $U_{\mu}$, the marginal of the visible nodes can be expressed as the exponential of a series involving all terms of the form $I^{(s)}_{k_1,k_2,\cdots,k_s} v_{k_1} \cdots v_{k_s}$ where  the interaction of order $s$ \cite{bulso2021restricted}:
\begin{equation}
    I^{(s)}_{k_1, k_2, \cdots, k_s } = \sum_{\mu=1}^M \sum_{p=0}^{s-1} (-1)^p \sum_{j_1 < \cdots < j_{s-p}=1}^{s} K\left( \sum_{i=1}^{s-p} w_{k_{j_l} \mu}\right)
    \label{eq:inter}
\end{equation}
and $K_{\mu}(x) = \log \mathbb{E}_{z_{\mu}}\exp(z_{\mu} x)$ is the cumulant generating function of $z_{\mu}\sim \rho_{\mu}(z)$ which is analytically related to $U_{\mu}$. Because of the characteristics of $K_{\mu}$, this structure implies that there is no RBM, endowed with a non-linear activation, able to express only terms up to a given order, for any choice of connections ${\bf w}$ \cite{bulso2021restricted}. 
These expressions and similar ones have been used to fit RBMs to interacting spin models, e.g. Ising models \cite{cossu2019machine, decelle2024inferring}.

{\bf Phase diagram and compositionality.} 
The formal equivalence between an RBM with linear hidden activation function and the Hopfield Model (HM) \cite{barra2012equivalence}, and its generalizations, have been exploited to investigate the properties of both networks \cite{marullo2020boltzmann} and to describe the phase diagram of RBMs \cite{AGLIARI2022232,PhysRevE.97.022310}. The HM is known to fail at retrieval when the number of stored patterns $p$ is such that $\alpha \equiv p/N > \alpha_c = 0.14$ \cite{amit1989}. Given the previously mentioned equivalence, the number of hidden units $M$ in the RBM plays the role of the number of patterns $p$ in the HM. The high storage load in the HM then corresponds to an RBM with a high number of hidden units and high number of parameters. The transition to no-retrieval above $\alpha_c$ in the HM, thus corresponds to the impairment of the RBM to learn as a result of over-fitting the data. What is remarkable here is that one concept, namely successful retrieval in attractor dynamics, is mapped to another, namely learning without over-fitting. This RBM-HM duality also relates two very different approaches to learning, supervised in HM and unsupervised in RBM, allowing insight gained from one to be applied to the other \cite{alemanno2023supervised}.\\
The analogy between the RBM with linear units and the HM is technically valid for binary weights in the RBM, as the HM is defined for binary units and patterns. Neurons and synaptic weights between them in the brain and in artificial neural networks are, however, not binary, and extensions of the HM to non-binary units and non-binary patterns have long been studied \cite{Treves_1990,kuhn1991statistical,Treves_1991,Bolle_1993,schonsberg2021efficiency}. It is thus a natural and fruitful step to study general RBMs and HMs, extending the results to beyond the binary and linear cases. This was the goal in several recent studies \cite{PhysRevE.96.042156, PhysRevE.97.022310, marullo2020boltzmann} that considered a generalized RBM, whose distribution is parameterized as
\begin{subequations}
\begin{align}
\begin{split}
p(\textbf{v},\textbf{z}) \propto \eta(\textbf{v})\ \rho(\textbf{z})\exp\left[\sum^{N,M}_{i,\mu=1} w_{i\mu} v_i z_{\mu}\right]
\label{eq:genRBM}
\end{split}\\
\begin{split}
p(\textbf{v}) = \sum_{\textbf{z}} p(\textbf{v},\textbf{z}) \propto \exp\left[\sum_{\mu}K\left(\sum_i w_{i\mu} v_i\right)\right]
\label{eq:genHopfield}
\end{split}
\end{align}
\label{eq:genRBMHOP}
\end{subequations}
\noindent where the functions $\eta$ and $\rho$ can be thought of as priors over the visible and hidden units. The marginal Eq. \eqref{eq:genHopfield} of Eq. \eqref{eq:genRBM} effectively describes a generalized HM with patterns $w_{i\mu}$ with $K$ having the same meaning as in Eq.\ \eqref{eq:inter}. One can interpolate between binary and Gaussian distributions using appropriate parametric forms of $\eta$ and $\rho$. The RBM-HM duality discussed above is then extended to more generalized versions of both architectures, relating transition to no-retrieval in one to over-fitting in the other; see \cite{marullo2020boltzmann} for a review.
The duality can prescribe a critical sample size after which the RBM is able to learn ``archetypes`` in datasets, a crucial property for successful generalization\cite{AGLIARI2022232}. It can also be used to study the learning of biased patterns \cite{AGLIARI2022126716} and yield a recipe for parameter initialization \cite{LEONELLI2021314}. Finally, we note that it is possible to prove, using Eq. \eqref{eq:inter}, that for linear hidden activation function and only in this case, all interactions beyond pairwise ($s\geq 3$) can be guaranteed to be equal to zero \cite{bulso2021restricted}. When such higher order interactions are present, the RBM may be related to the so called Dense Associative Networks that express exponential storage capacity \cite{krotov2023new,lucibello2024exponential}.\\
A {\it compositional phase} is a phase where the visible configurations are generated from a composition of a large number, $L$, of strongly activated hidden units ($1 \ll L \ll M$). Active hidden units corresponding to a given visible configuration provide a simple representation of that configuration. The analysis of the ground states of Eq.\ \eqref{eq:pdef} for an ensemble of RBMs with random i.i.d. weights and ReLU hidden activation function demonstrates the existence of such a phase in RBMs \cite{PhysRevLett.118.138301}, e.g. when the weight matrix is sparse. Similar results are also shown for RBMs with binary hidden and visible nodes and extending the analysis to the more realistic case where the weights in the RBM ensemble are not i.i.d. but drawn from more realistic distributions \cite{Decelle:2018ab}.\\
Having discussed recent theoretical work emphasizing the role of activation functions in RBMs, below we turn to reviewing a range of applications, demonstrating this role in modeling data from different biological systems. This will show how RBMs offer simple, interpretable models for uncovering features in these data.

{\bf Applications to biological data.} 
\setlength{\parskip}{0pt}
Despite their simplicity, RBMs have proven useful in many applications, ranging from representing the states of quantum many body system \cite{carleo2017solving,hartmann2019neural} to biological data analysis. Our focus here will be on applications in neural data analysis, protein families and immunological data. Given the binary nature of spiking in neurons, application to neural data often involves RBMs with binary units and sigmoid activation function. However, as we will discuss, this needs to be changed when RBMs are applied to protein families and immunological data.

{\it Neural Data.} RBMs have been used to model correlations within micro-columns in the cat visual cortex \cite{10.1371/journal.pcbi.1003684}. Compared to pairwise Ising models, RBMs exhibit better performance in modeling data statistics, highlighting the importance of high order correlations in the cortex \cite{olsen2024quality}. Increasing the scale of the dataset by several orders of magnitude, Van der Plas et al. \cite{van2023neural} trained RBMs in the compositional phase to the spontaneous activity of larval zebra fish, finding $\sim 200$ anatomically localized functionally relevant neural assemblies. Applied to the activity of retinal ganglion cells \cite{volpi2020modeling}, RBMs have been used to construct metrics for neural responses that can help associate or discriminate sensory stimuli \cite{gardella2018}. 
Extensions of RBM involving dynamics, dubbed Temporal RBM (TRBM) \cite{sutskever2008recurrent}, have also been used to construct such metrics taking into account temporal aspects of neural dynamics \cite{volpi2020modeling, gardella2018}. TRBMs involve a sequence of RBMs, where the state of an RBM at time $t$ depends on the state at time $t-1$, that is
\begin{equation}
    p(\textbf{v}_t,\textbf{z}_t| \textbf{z}_{t-1}) = \frac{e^{\left[ \textbf{v}_t^\top \textbf{b} + \textbf{v}_t^\top \textbf{W}\textbf{z}_t + \textbf{z}_t^\top\left(\textbf{c} +  \textbf{W}^{'}\textbf{z}_{t-1}\right) \right]}}{Z(\textbf{z}_{t-1})}
\end{equation}
where $\textbf{W}^{'}$ is the weight matrix connecting the hidden biases at time $t-1$ ($\textbf{c}_{t-1}$) to those at time $t$ ($\textbf{c}_t$). 
\begin{figure*}
\centering
\includegraphics[width=18cm]{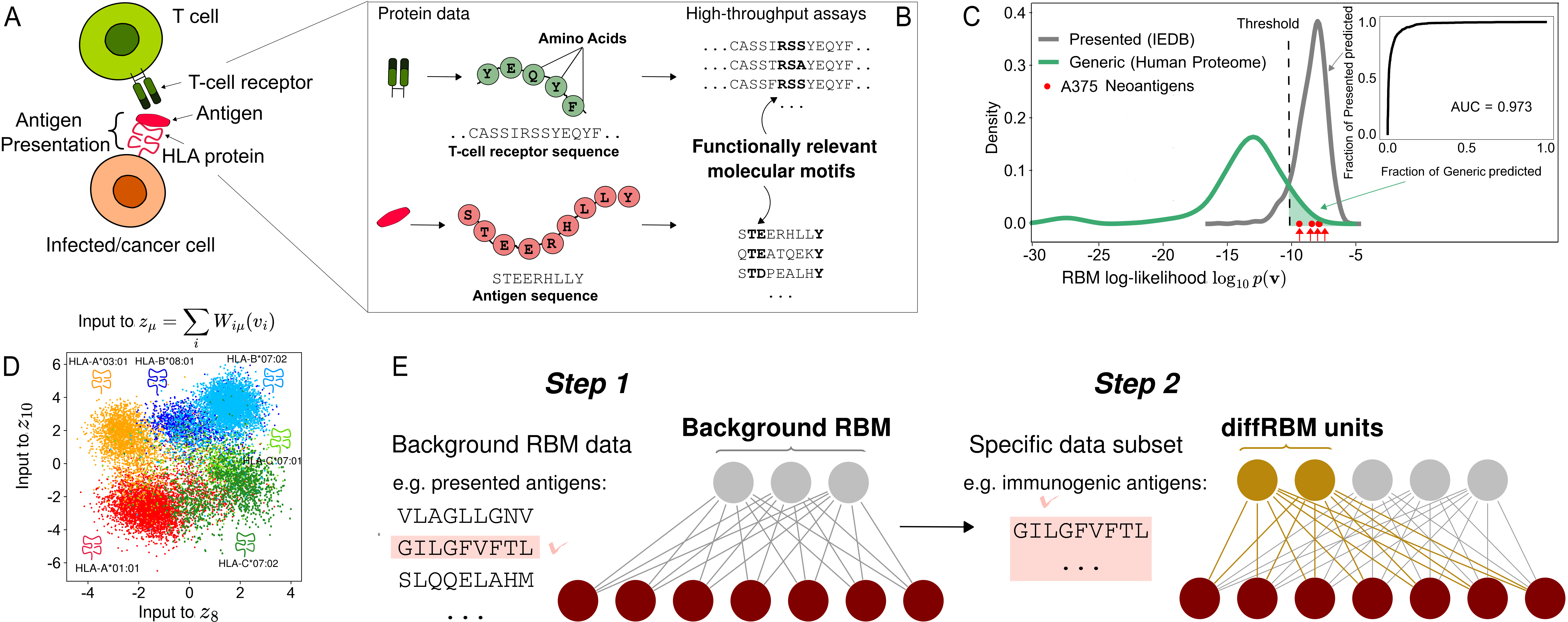}
\caption{\textbf{RBM applications to protein data in immunology.} A: T cells recognize cancer and infected cells via the binding of T-cell receptors to antigens presented on the cell's surface by HLA proteins. B: High-throughput experimental and sequencing platforms yield large datasets sampling the proteins involved in immune recognition (T-cell receptors, antigens). C: The RBM log-likelihood gives a probabilistic score that can discriminate functional from non-functional proteins: in this example, antigens presented by a specific HLA (including cancer-related antigens, in red) from generic non-presentable protein fragments. D: The RBM latent representations group protein sequences into functional subfamilies: in this example, antigens presented by different HLA types. Panels C and D were adapted from \cite{bravi2021}. E: 2-step learning of diffRBM \cite{bravi2023}, in this example applied to modeling antigen immunogenicity.}
\label{fig:biological}
\end{figure*}
TRBM is difficult to train, but a similar dynamical variant of RBM, the Recurrent Temporal RBM (RTRBM) can be trained efficiently and exactly \cite{sutskever2008recurrent}. When applied to whole-brain larval zebra fish recordings, the resulting RTRBM model extends upon the identified neural assemblies by additionally capturing their temporal dependencies \cite{quiroz2024recurrent}. The application of this kind of machine to neural data is a promising and exciting direction, where efficient learning, interpretability and temporal structure detection come together.

{\it Protein Families.} RBMs have been applied to protein data for performing structural prediction, computational design, functional characterization and classification \cite{tubiana2019, shimagaki2019, malbranke2021, tubiana2023, malbranke2023, mauri2023a, decelle2023}. Although in neural data analysis units in an RBM are often taken to be binary with sigmoid activation function, in protein applications, one needs to go beyond these choices. This is because a protein is a sequence of amino acids, with 20 different amino acids existing in total. To use RBMs for modeling these sequences, one thus needs to consider Potts units: each $v_i$ then stands for the amino acid at position $i$ along the sequence and takes discrete values within an alphabet of size 20.\\
In different species, the same functional protein may be represented by a somewhat different amino-acid sequence, reflecting evolutionary changes in the protein. Protein datasets usually contain such differing sequences of the same protein from different species. The statistical properties, e.g., the frequency of an amino acid at a certain position and their co-occurrence at different positions, are then taken to reflect functional and structural constraints on the protein's biochemical composition. Capturing the statistical relationships between amino acid usage at different positions, RBMs can be used to find and study such constraints. \\
Tubiana et al.~\cite{tubiana2019} showed that the RBM weights learned from protein sequences can recover contacts in the 3D structure in a manner akin to the popular technique Direct Coupling Analysis \cite{morcos2011}, detecting a variety of structurally and functionally relevant sequence motifs. By parametrizing a multi-modal probability distribution over protein sequences, the RBM weights reflect multiple sequence patterns that correspond to the biochemical properties of distinct functional subfamilies, differing in terms of activity or ligand-binding specificity. The RBM is able to group the modeled proteins into such functional subfamilies in the low-dimensional latent-space representation of the sequence data obtained by projecting them onto the RBM weights. Two aspects bench-marked in \cite{tubiana2019} to expand the capabilities of the RBM architecture for protein modeling included expressive power and interpretability. Emphasizing the role of hidden units' activation function, it was shown that a high expressive power particularly effective for learning high-order correlations is conferred by
\begin{equation}
U_{\mu}(z) = \frac{1}{2}\gamma_{\mu,+} z_{+}^2 + \frac{1}{2}\gamma_{\mu,-} z_{-}^2 + \theta_{\mu,+} z_{+} + \theta_{\mu,-} z_{-}
\label{eq:dReLU}
\end{equation} 
where $z_{+}=\max(z,0)$, $z_{-}=\min(z,0)$, and $\gamma_{\mu,+}$, $\gamma_{\mu,-}$, $\theta_{\mu,+}$, $\theta_{\mu,-}$ are learned from data. This potential represents a double Rectified Linear activation function.

{\it Protein design.} Putative functional protein sequences can be generated by using RBMs trained on data as generative model to sample from \cite{malbranke2021, tubiana2023, malbranke2023,carbone2023}. RBM can therefore also be useful for computational protein design. Experimental tests demonstrated that these ensembles contain truly functional sequences, which can be identified even more efficiently when the RBM-generated sequences are further screened and filtered for their expected biophysical properties, e.g., stability, solubility, docking energies, by means of complementary computational methods \cite{tubiana2023,malbranke2023}. \\
An advantage of the RBM that remains to be fully explored is the fact that its latent-space representations can suggest how to tune the sampling to design sequences belonging to specific functional subfamilies \cite{tubiana2019} or to generate viable intermediates between subfamilies \cite{mauri2023a}. The generative and pattern detection properties of the RBM have led to their successful employment for molecular design purposes also with RNA sequences \cite{gioacchino2022, fernandez-de-cossio-diaz2024}.

{\it Immunology}. RBMs have been used in immunological data analysis \cite{bravi2021, bravi2021a, luksza2022, bravi2023}, where large-scale experimental datasets of proteins involved in immune responses are available. These works have focused on the protein-protein binding processes that trigger the response to infections and cancer progression by killer T-cells in human. T-cells are one of the main cell types of the adaptive immune system. 
In order to respond, the presence of an infected or a malignant cell must be signaled to T-cells via a step called ``antigen presentation'' (Fig. \ref{fig:biological}A): a protein called Human Leukocyte Antigen (HLA) binds to antigens, short protein fragments from the pathogen or harboring cancer-related alterations, forming the HLA-antigen complex, and in this way it exposes (``presents'') antigens to T-cells on the surface of the infected or malignant cell.\\
T-cells possess surface proteins (T-cell receptors) that can bind specifically to a given HLA-antigen complex, activating their response to a pathology. RBMs trained on experimental data can be used for modeling the probability distribution over the sequence in HLA-presented antigens \cite{bravi2021}, and T-cell receptors probed for their response to selected antigen targets \cite{bravi2021a}. The trained model embeds the distinctive amino acid usage that characterizes antigens presented by certain HLAs and T-cell receptors specific to a given antigen and can be used to assign probabilities to new amino acid sequences. These probabilities can serve as a score to discriminate antigens presentable by a certain set of HLAs from generic non-presentable short protein fragments (Fig. \ref{fig:biological}C), as well as antigen-specific from unspecific T-cell receptors, with a performance that compares favorably to state-of-the-art methods.\\ 
As shown in Fig. \ref{fig:biological}D, the representation of the data by the RBM hidden layer can be used to cluster antigens and T-cell receptors. Such clustering reflects the shared amino acid motifs that contribute to determine the binding specificity of antigens towards their presenting HLA molecule and of T-cell receptors towards their cognate antigen-HLA complex. Using such clusters, one can predict the antigen binding specificity to the HLA with high accuracy \cite{bravi2021}.\\
Predictions on antigen presentation and T-cell response specificity are extremely relevant to vaccine and therapy design, as they can accelerate and guide the selection of vaccine targets \cite{bravi2024}, T-cell engineering, and the discovery of cancer-specific antigens for the development of anti-cancer immunotherapies \cite{roudko2020}. RBM probabilistic scores for T-cell receptors response have been exploited to detect statistical signatures of immune activity in patients \cite{luksza2022}, thus helping to rationalize the role of the immune system in cancer evolution and the survival of patients.

{\it Differential RBM}. The strength of interpretable yet powerful machine learning approaches like RBM in quantitative immunology is linked not only to prediction, but also to discovery and better characterization of the molecular properties that enable effective immune responses. To this end, Bravi et al.~\cite{bravi2023} proposed a novel strategy of ``differential'' learning within the RBM (diffRBM), and applied it to immune protein data to disentangle amino-acid motifs underlying on the one hand antigen immunogenicity (the ability of certain antigens to trigger potent responses), and on the other hand the specificity of binding of T-cell receptors to an antigen.\\
The diffRBM strategy (Fig. \ref{fig:biological}E) is an example of transfer learning, as it consists of the following two steps: 1. learning a first RBM model (background RBM) describing baseline constraints exhibited by the general type of data under consideration (e.g., constraints on amino acid compositions shared by generic, viable antigens or T-cell receptors for the application to immune proteins); 2. transferring background RBM to the subsequent task of learning the statistical distribution of the specific subsets of data that are the final modeling target (e.g., immunogenic antigens, or antigen-specific T-cell receptors). In step 2, a few hidden units (diffRBM units) are added to the RBM from step 1, learning only their corresponding parameters and weights while keeping all the other parameters fixed: the diffRBM units are thus forced to learn the statistical differences in these specific subsets of sequences, e.g., distinctive molecular motifs in immune protein data. The differential mode of learning facilitates the inspection of such motifs by concentrating them on a few units and allows the use of them for the prediction and identification of the TCR-antigen-HLA binding sites \cite{bravi2023}. The strategy of differential learning by diffRBM is more general than the specific application analyzed in \cite{bravi2023}, it has thus the potential to be further harnessed in the future to learn selected or enriched statistical features in biological data.

\begin{figure*}
\includegraphics[width=1\linewidth]{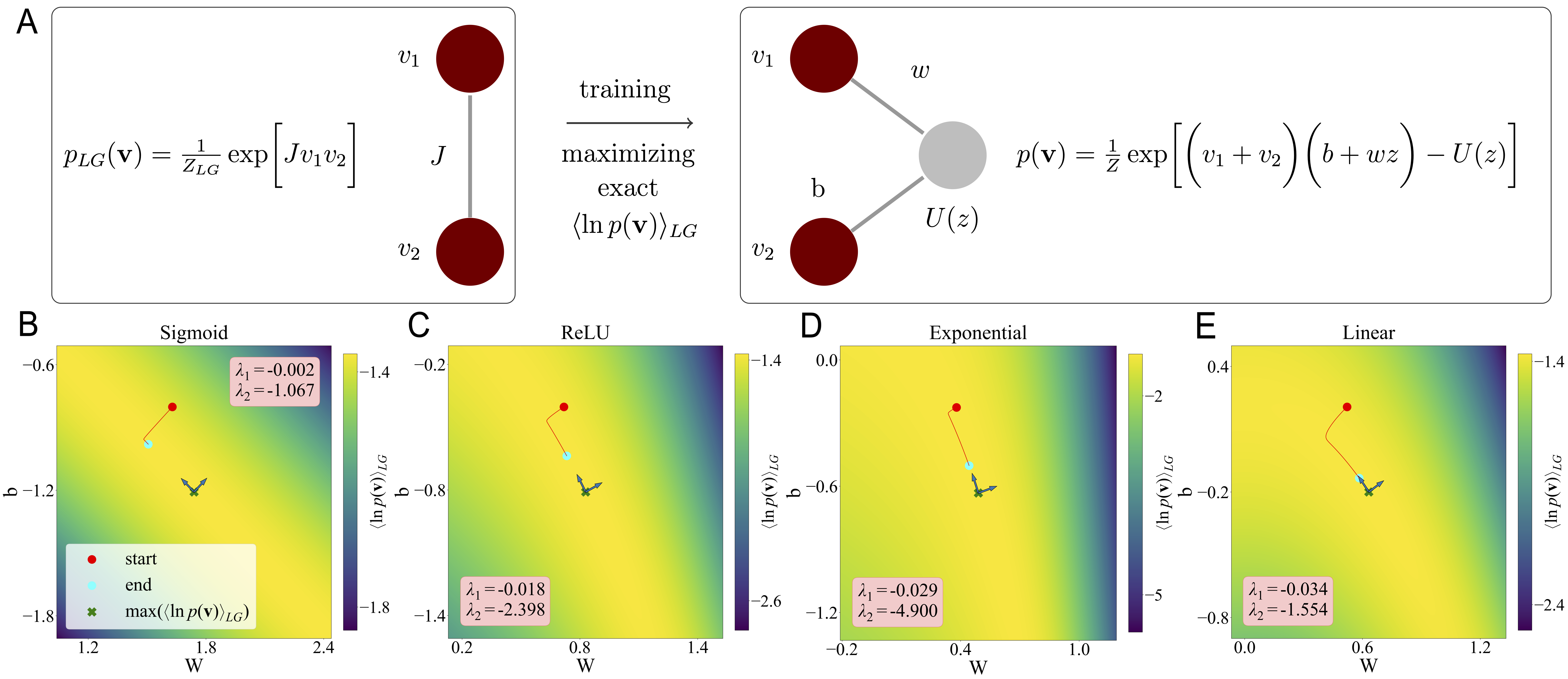}
\caption{\textbf{Effect of hidden units non-linearity on learning in simple RBMs.} In the simple case of panel A, RBMs with $N=2$, and $M=1$ and different hidden potentials $U(z)$ are trained by maximizing the log-likelihood for data from $p_{LG}({\bf v})$ exactly. $\langle \cdot \rangle_{LG}$. B-E: The log-likelihood is computed exactly by enumerating the averages over $p_{LG}({\bf v})$ in the $(w,b)$ plane for 4 different potentials $U(z)$ corresponding to (B) step, (C) ReLU, (D) exponential, and (E) linear hidden activation functions. Inverting Eq.\eqref{eq:inter}, the ($w,b$) values corresponding to $h=0, J=0.4$ are denoted by green crosses. These also correspond to the maximum likelihood values. The arrows represent the eigenvectors of the Hessian of the likelihood computed at its maximum, with eigenvalues reported in the legend. Although $\lambda_1$ is always close to zero, the difference in its amplitude between different activation function leads to a substantial difference in speed of convergence. The learning trajectory with the same number of gradient descent steps, learning rate and initial condition (red circle) are shown, ending at the cyan circle.}
\label{fig:logl}
\end{figure*} 
\setlength{\parskip}{6pt}

{\bf Discussion.} Statistical physics approaches to data analysis have a long history \cite{cocco2022statistical}. What we reviewed here shows that among the data analysis models studied through the lens of statistical physics, RBMs play a special role. Firstly, having comparatively lower parameters than deeper architectures, they offer remarkable benefits in biological applications where data is large but often not large enough to train deeper networks. Despite their simplicity, the applications reviewed here show that they can still offer important insights into biological systems, demonstrating that they are powerful in capturing non-trivial relationships in data. Second, this simplicity also makes them interpretable, another advantage for biological data analysis. Finally, they are amenable to analytical inspection for problems of general importance to neural network research. In what follows, we list a number of such problems.
\setlength{\parskip}{0pt}

{\it Optimal activation function.} Neural networks used in practice or in theoretical analyses, often employ one of few forms of activation function, e.g. step, ReLu, dReLu. Making this choice is often based on heuristics. RBMs can provide a platform for studying whether one can make such choices in a systematic way, e.g., based on general knowledge on the properties of the data. In the case of associative networks of memory, recent work shows that the choice of activation function makes a marked difference in the performance of the network, in particular when storing sparse activity patterns \cite{schonsberg2021efficiency}. Analyzing RBMs with different activation functions, e.g. as in \cite{bulso2021restricted}, can be a fruitful approach to understanding how activation functions can influence the range of distributions that a given RBM can represent. When combined with a better understanding of learning dynamics in RBMs (see below), this may offer a systematic approach to choosing appropriate activation functions for efficient learning in RBMs and beyond.

{\it Learning Dynamics and Training}. Studying learning dynamics in neural networks using statistical physics has a long history; see, e.g.,~\cite{saad1995line,saad1995exact,saad1998learning,goldt2019dynamics,loureiro2021learning}. Most of these studies, however, have focused on the case of supervised, student-teacher learning, while analytical approaches to studying learning dynamics in unsupervised stochastic machines are comparatively scarce. RBMs can be an ideal model for such analyses. Existing work, however, demonstrate that understanding the learning dynamics can yield insight into the operational phase of RBMs trained on real data \cite{decelle2017spectral,Decelle:2018ab} and to produce training approaches that improve the performance of RBMs \cite{decelle2021equilibrium}. Given this, an important question is how the form of the activation function, encoded in the potential $U_{\mu}$ in Eq.\ \eqref{eq:pdef}, affects convergence to local or global maxima of likelihood. Fig. \ref{fig:logl} shows a simple toy example that demonstrates that the curvature of the likelihood function and the convergence time vary considerably between different activation functions, although all of them may eventually converge to the optimal solution. Understanding this dynamics and choosing appropriate activation functions in a given task can help design powerful generative RBMs that can be trained in a reasonable time. It would also be interesting to see how activations functions influence phase transitions observed during learning for RBMs with binary units \cite{bachtis2024cascade}, and the influence this may have for successful training. Insight gained by these analyses in the case of RBMs can act as a stepping stone for designing more efficient unsupervised learning algorithm in deeper networks.

{\it Overparametrization and noisy learning}. An important feature of current deep neural network architectures is the overparametrization of the system. This feature is also probably shared by neural networks in the brain. However, an important difference between the two is that learning in the brain is noisy: biological learning rules cannot tune parameters to desired values with high precision, and the tuned values will fluctuate over time even when they should not. In artificial neural networks, one has the opposite scenario: parameters can be tuned with high precision and kept to those values. Again, here RBMs can be an excellent first step for understanding the effects of noisy unsupervised learning in overparametrized networks to gain better insight about unsupervised learning in biological neural networks. \acknowledgments \bibliography{refs} \end{document}